\documentclass[12pt,preprint]{aastex}
\usepackage{graphicx}
\slugcomment{Accepted by The Astrophysical Journal 9/01}
\shorttitle{FUV EXTINCTION CURVE}
\shortauthors{SASSEEN ET AL.}

\begin{document}
\title{A New Measurement of the Average FUV Extinction Curve}
\author{T.P. Sasseen\altaffilmark{1}, M. Hurwitz\altaffilmark{2}, W. V. Dixon\altaffilmark{2,3} \& S. Airieau\altaffilmark{2,4}}
\altaffiltext{1}{Dept. of Physics, U. C. Santa Barbara, Santa Barbara, CA 93106, USA}
\altaffiltext{2}{Space Sciences Laboratory, U. C. Berkeley, Berkeley, CA 94720, USA}
\altaffiltext{3}{Current address: Department of Physics and Astronomy,
The Johns Hopkins University, Baltimore, MD 21218, USA}
\altaffiltext{4}{Current address: San Diego State University, Astronomy Dept., 5500 Campanile Drive, San Diego, CA 92182, USA}

\begin{abstract}
We have measured the extinction curve in the far-ultraviolet
wavelength region of (900 -- 1200 \AA\/) using spectra obtained 
with the Berkeley EUV/FUV spectrometer during the ORFEUS-I and
the ORFEUS-II missions in 1993 and 1996.  From the complete sample
of early-type stars observed during these missions, we have 
selected pairs of stars with the
same spectral type but different reddenings
to measure the differential FUV extinction. 
We model the effects of molecular hydrogen absorption
and exclude affected regions of the spectrum 
to determine the extinction from dust alone.  
We minimize errors from inaccuracies in the cataloged spectral types 
of the stars by making our own determinations of spectral types 
based on their IUE spectra.   We find substantial scatter in the curves of individual
star pairs and present a detailed examination of the uncertainties
and their effects on each extinction curve.
We find that, given the potentially large uncertainties inherent in 
using the pair method at FUV wavelengths,  a careful
analysis of measurement uncertainties is critical to assessing the true
dust extinction.  We present a new measurement of the average 
far-ultraviolet extinction curve to the Lyman limit;
our new measurement is consistent with an extrapolation of the 
standard extinction curve of Savage \& Mathis (1979). 
\end{abstract}


\section{INTRODUCTION}

Dust in the interstellar medium is the dominant continuum absorber 
over wavelengths from the far ultraviolet (FUV) to the infrared.  
An understanding of dust extinction is important not only for the correct
interpretation of astronomical flux measurements, but also because
it can be used to investigate the composition, scattering and absorption 
properties, and history of the dust itself.  One of the 
least-studied regions of the dust extinction
spectrum is the FUV, precisely where the extinction is greatest.
In this study, we have used data from the Berkeley spectrometer on
the ORFEUS telescope taken during 
its flights in 1993 and 1996 (Hurwitz et al. 1998)
to study dust extinction in the FUV.  Our data cover the region 900 -- 1200 \AA,
or 8.33 to 11.11 $\mu$m$^{-1}$.  A preliminary study, based only
on data from the first flight, was presented by Sasseen et al. (1996).  

Measurements of the Galactic absorption curve from 1 micron to 1110 \AA\/
are summarized by Savage \& Mathis (1979).  Aside from atomic and molecular lines, 
such as the H$_2$ bands beginning shortward of 1120 \AA, the
dust absorption curve has been shown to vary smoothly
from the FUV to near-infrared wavelengths by Cardelli, Clayton
\& Mathis (1989), based on the measurements of Fitzpatrick \& Massa (1986, 1988).  
Although their work does not include the wavelengths measured here,
this finding implies a continuous distribution of dust properties over
the optical and ultraviolet wavelength range.  Cardelli et al. 
parameterize the extinction curve in terms
of $R_v$, the ratio of $ A_V$ to $E(B-V)$, and express the 
extinction curve based on analytical fits to the data.
The 2175 \AA\  bump, attributable to graphite dust, is studied by 
Fitzpatrick \& Massa (1986), and the extension of the extinction curve to
shorter wavelengths by Fitzpatrick \& Massa (1988).

Measurements of extinction curves in the FUV (here 
$ 912 - 1216$ \AA) have been made by
Longo et al. (1989) and Snow, Allen \& Polidan (1990) 
using Voyager 1 and 2, IUE,
and TD-1 data, and by Green et al. (1992) with a rocket-borne
experiment.  Green et al. measure the extinction towards $\rho$ Oph, a region
which is known to show anomalous dust extinction, attributable
to an excess of large dust grains and a deficit of small
grains.  Green et al. find that the FUV extinction curve 
for $\rho$ Oph is consistent with a simple extrapolation from
the UV curve measured at larger wavelengths by Fitzpatrick
\& Massa (1990), but cannot be fit by the standard grain 
composition model (Draine \& Lee 1984, 1987).
FUV extinction curves in the local spiral arm were also measured by Buss et al. (1994)
using HUT and IUE spectra.  These authors explore extinction
in a number of different Galactic environments
and discuss the effects of environment on grain size.
However, they are hampered by inexact spectral matches and
data obtained with different instruments that have lower resolution
than ORFEUS.  The lower resolution makes it more difficult to 
differentiate the real continuum from absorption by H$_2$ and other
interstellar species.  

Recent work on the nature of FUV extinction curves in the 
Large Magellanic cloud is presented in Misselt, Clayton \& Gordon (1999),  
and for the Small Magellanic Cloud by Gordon, \& Clayton (1998).
A review of other extragalactic measurements is given by Fitzpatrick (1989).  
Misselt et al. find the general properties of 
LMC extinction similar to regions of the Galaxy, but find that the correlation between
environment and extinction to be different than for the Milky Way. 
These authors also find distinct differences between the general LMC
extinction and the extinction within 30 Dor and the LMC 2 supergiant shell.  
These studies are interesting because one reason to study 
extinction is to understand the detailed interrelationship between dust, 
gas densities, H$_2$ formation rates, the effects of ambient radiation, 
metallicity and history on the composition of the ISM.  
External galaxies provide a broader parameter space for investigating
these effects.  

Given observations of the dust extinction curve, researchers frequently
match the observed extinction with the predictions from a model based
on the dust size and composition. The widely-used
Draine \& Lee (1984) model (with additional corrections by
Draine \& Lee 1987), uses grains of various sizes and composition, along 
with measured and calculated values for their scattering 
and absorption properties, to predict an extinction curve.
Weingartner \& Draine (2001) are able to reproduce many properties of
Magellanic Cloud and Milky Way UV -- IR extinction curves by modeling the size distribution
of carbonaceous and silicate grains.  
The inclusion of new populations of absorbers, such as polycyclic aromatic
hydrocarbons (Puget \& L\'{e}ger 1989) and amorphous carbon (Colangeli et 
al. 1995) have been shown to be necessary to explain certain 
absorption features from the UV to the IR.  

Peculiar extinction curves, that is, curves that deviate substantially
in shape or magnitude from the average extinction curve, have been 
noted in the UV and studied by several authors (Mathis \& Cardelli, 1992, 
Savage \& Mathis 1979; Massa, Savage \& Fitzpatrick 1983). 
Despite differences in the extinction curves seen in various
Galactic look directions, astronomers frequently employ a "mean extinction curve"
to predict or correct for the effects of Galactic extinction.  In part
because of this practice and its evident utility, it is worthwhile to 
measure and establish an average extinction curve.   

In this paper, we measure and present the average extinction curve from 900 -- 1200 \AA\/, 
appropriate for the diffuse intersteller medium ($R_v = 3.1$), 
derived from stellar observations obtained
with the Berkeley spectrometer during the ORFEUS I and ORFEUS II missions. 
The ORFEUS data set is well suited to an investigation of FUV extinction owing
to its high resolution and the large number of stars observed during the two missions.  
We present here the first detailed study of the FUV extinction curve based
on this new data set.  
We describe the data selection and winnowing and how our
extinction curve is derived.  We attempt to minimize errors in the 
extinction curve
owing to mis-matched spectral types and the presence of H$_2$ absorption
through much of the band.  In the first case, we have made our own 
determination of the spectral types of stars based on IUE spectra.  In
the second, we have used the models of 
Dixon, Hurwitz \&\/ Bowyer (1998) to identify
regions of significant H$_2$ absorption and exclude
these wavelengths from our analysis of the dust absorption.  Of course, 
when correcting measured stellar fluxes to find their intrinsic
brightness, the effect of H$_2$ absorption should also be included. 
We then analyze the uncertainty in individual extinction curves and
select those with the lowest uncertainty to use in our determination
of the average curve.  Finally, we discuss how the new curve
compares with previous measurements.  

In a subsequent paper, Sasseen et. al (in preparation, hereafter paper II), 
we will use the individual 
extinction curves to infer the effects of local environment and dust
processing in these environments.  

\section{THE DATA AND METHOD}

The pair method of measuring extinction curves relies
on observing carefully-selected pairs
of stars with very similar spectral type but different amounts
of reddening.  After correcting for intrinsic differences in magnitude and
absorption by H$_2$, further differences
in the flux level of the two spectra are attributed to dust.
Following Savage \& Mathis (1979) and Green et al. (1992), we 
derive the extinction curve from flux ratio via 
\begin{equation}
E(\lambda - V) = -2.5\: \log \left(\frac{F_{red}(\lambda)}{c(\lambda) F_{st}(\lambda)}\right) + (V_{st} - E(B-V)_{st}R_{v} - V_{red}) .
\label{eq:green}
\end{equation}
Here, $V$ is the visual magnitude,
$E(B-V)_{st}$ is the color excess of the standard star,
$R_{v}$ is the ratio
of total extinction $A_{v}$ to $E(B-V)$, $F_{red}$ and
$F_{st}$ are the flux from the reddened and standard stars, and $c(\lambda)$
is the dereddening correction applied to the standard star.  
We use $R_v = 3.1$ throughout this paper, appropriate for 
the diffuse ISM.  We show below that 
minor deviations of $R_v$ from this value have essentially 
no impact on the final measurement of the extinction curve.  

\subsection{Establishing Spectral Type and Basic Astronomical Data}

During the ORFEUS I and II missions, a total of 41 stars were observed
that have stellar types earlier than B4, a rough cutoff such that 
stars with sufficient extinction for a measurement have well-
determined continuum fluxes down to the the Lyman limit.  
We obtained basic astronomical data 
for this initial set of stars and applied a number of selection criteria
to achieve a final sample that is best suited for determining the extinction 
curve. 

We obtained position and variability data from references within the SIMBAD 
database, but found it necessary to verify the spectral types of the stars ourselves 
to achieve sufficient accuracy in $E(B-V).$  
We made our spectral determinations by comparing archival IUE 
data of ORFEUS stars for those stars where available
with the standard UV stellar
spectra of Rountree \& Sonneborn (1993).  As a number of authors
have discussed, (e.g. Cardelli et al. 1992), it is most important 
when studying extinction that stellar pairs exhibit nearly identical 
photospheric characteristics in the wavelength range of comparison, 
rather than necessarily conforming exactly
to a particular MKS spectral class.  Absolute spectral classes determined
from UV data may differ slightly from those derived solely from
optical data, but these differences are generally minor (Rountree \& Sonneborn 1991).  
For the purposes of this paper, in which we are specifically
trying to match UV fluxes, we are justified in performing spectral matching
from UV data alone.  We used the procedure
of Rountree \& Sonneborn (1991), making 
a large-format, normalized hardcopy plot of each spectrum and 
visually comparing it with the standard spectra.  Photospheric 
and wind-line equivalent widths are the main diagnostic parameters,
with the former taking precedence.  

We show in Table~\ref{tab:basic} the spectral types we derive. 
We estimate our overall uncertainty to be 0.3 in spectral
type and about half a luminosity class.  There are significant differences
between the spectral types we determined and previously published 
values in several cases.  We use the spectral types from our study when available to 
group stars by spectral type.  A few stars did not have archival IUE observations
suitable for us to make spectral determinations, so we adopted the spectral
type from the listed references for these stars.  

We use "mean UBV" V magnitudes and \bv colors from the General Catalog
of Photometric Data (Mermilliod et al. 1997, hereafter GCPD).   The mean values listed in
this catalog are derived typically from an appropriate weighted average of 
several photometric measurements the authors deem reliable.  
To determine $E(B-V)$ for each star, we used \bv from the Catalog
and $(B-V)_0$ for each spectral type and luminosity class from 
Fitzpatrick \& Massa (1990), Fitzgerald (1970), and 
Mihalas \& Binney (1981).  The basic parameters we adopt for each 
of the stars are presented in Table~\ref{tab:basic}.  

From this list, we further eliminated a number stars from further
study because their spectra showed peculiarities; there
was not a suitable low- and high-extinction star with the same spectral
type; the basic astronomical data were lacking or suspect; or 
the star was bright enough to cause gain sag in the ORFEUS detector
(flux at 1050 \AA\/ greater than 1.1 
$\times 10^{-10}$ ergs s$^{-1}$ cm$^{-2}$ \AA$^{-1}$, Hurwitz et al. 1998),  
leading to uncertain spectral flux. 
We also eliminated known variable stars since we do not have 
simultaneous V measurements of the stars measured by ORFEUS.  
The final sample consists of 18 stellar pairings, 
presented in Table ~\ref{tab:fits}; 
most pairs have identical spectral type, if not luminosity class.  
Cardelli, Sembach \& Mathis (1992) show that giants and supergiants,
some of which we include in our sample, can be suitable for measuring UV 
extinction curves, but caution that spectral type and luminosity class should
be matched in a pair.  We evaluate the effect of including these stars in our
sample below.  

One obvious aspect of the spectra is the large number of absorption
lines, primarily due to H and H$_2$,  that make
it difficult to identify a continuum level and hence determine
the continuum extinction due to dust alone.  
With spectral resolution of 0.3 \AA, 
ORFEUS has sufficient resolution to separate
H$_2$ absorption lines from the continuum.  
We use the models of Dixon et al. (1998) to indicate where H and H$_2$ absorption 
is present.  The models use inputs of column density, the Dopper $b$ parameter
and the relative velocity of a given
species to compute a transmission spectrum.  
Regions in the spectrum of a highly-extincted star that are absorbed by 
more than 4\% by H$_2$ absorption lines are not used in calculating the continuum level in
any of the spectra.  In addition, the model results show that for the five most 
heavily extincted stars, even this correction is not sufficient to completely 
remove the effects of absorption by H$_2$.   An additional correction for stars 
whose continuum measurement would be affected by more than 3\% 
(HD 103779, HD 109399, HD 113012, HD 37903, HD 99857) was applied by
a smooth fit to the least absorbed parts of the spectrum over the wavelength 
range 900 - 1150 \AA.  
The average correction applied for these stars over this range is 5\%.  
The detailed results of fitting the stellar
continua are reported in Dixon et al. (2001).  

To evaluate whether there is any residual ISM line absorption not
removed by our modeling procedure, we used the measurements of Morton (1978) 
to identify where significant absorption is present in our band.  Morton uses
high resolution {\em Copernicus} spectral measurements of Zeta Puppis to measure 
the equivalent widths of detected ISM lines.  Of the 25 absorption lines Morton 
measures with equivalent widths greater than 100 m\AA\/ (an arbitrary cutoff) 
between 920 and 1190 \AA\/, 9 are
specifically in the H$_2$ model, 13 are in regions excluded because they are 
close to H$_2$ absorption, and only three are within our selected continuum regions. 
The small equivalent widths of these three remaining lines in even the most heavily absorbed 
stars in our study affects our overall continuum level placement by less than 1\%, 
a negligible amount compared with our other uncertainties.

\begin{figure}
\plotone{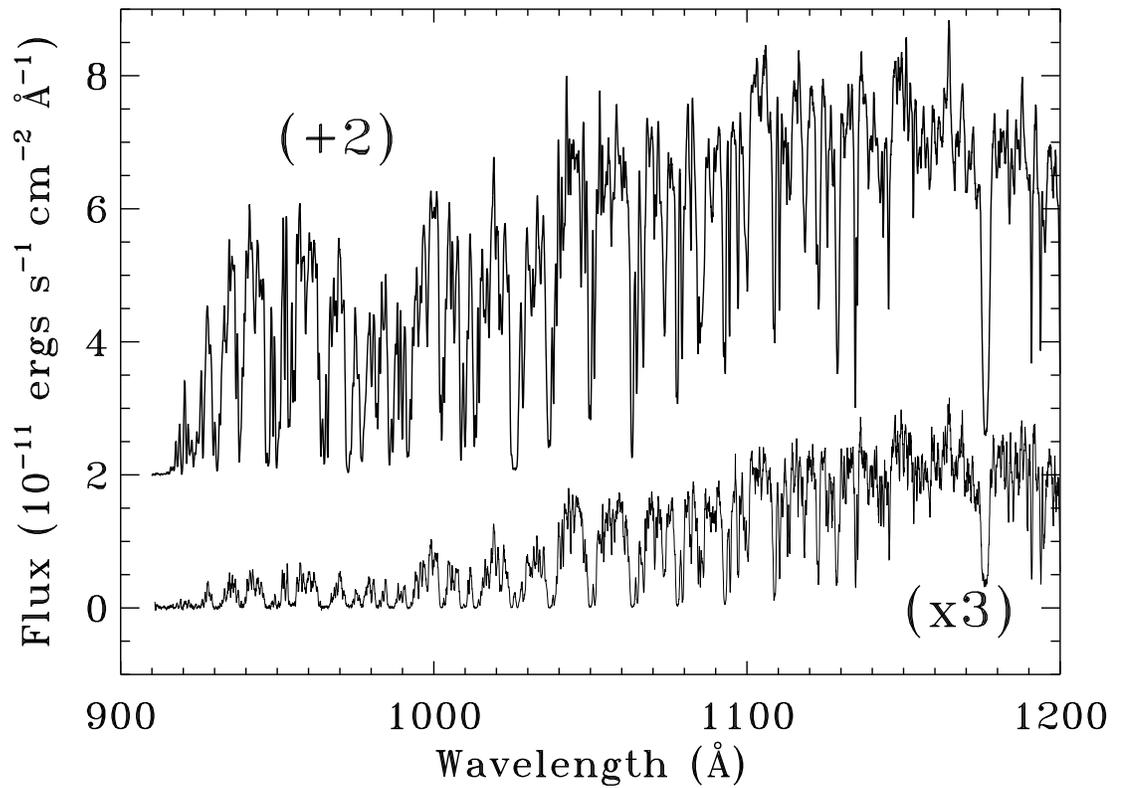}
\caption{The FUV spectra of HD 186994 (upper) and HD 113012 
(lower) taken with the Berkeley spectrometer on the ORFEUS telescope. 
Significant H$_2$ absorption is evident in the both spectra; we
model H$_2$ in the spectra and exclude regions of the spectrum that
show significant absorption by H$_2$.  \label{fig:spec} }
\end{figure}

\begin{figure}
\plotone{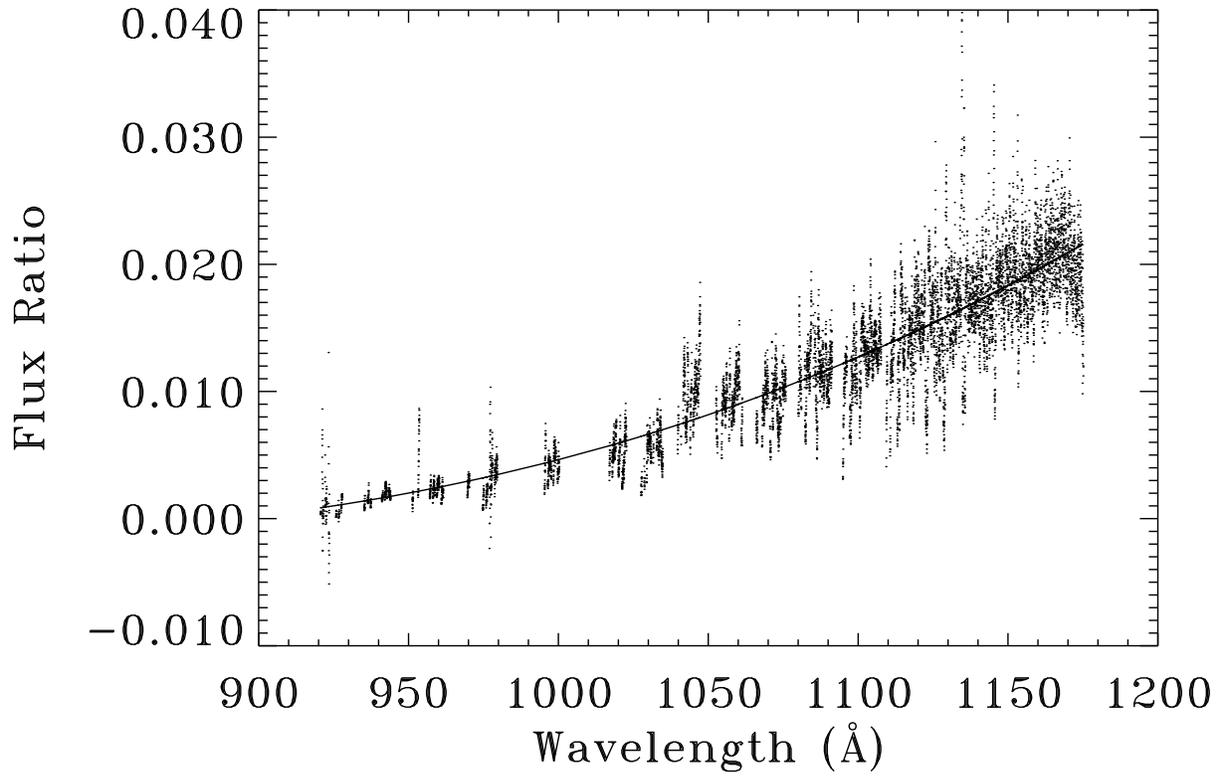}
\caption{The ratio of the two spectra shown in Fig. 1
and the polynomial fit that characterizes the curve.  Significant flux
is detected in the two stars to 920 \AA\/.  
\label{fig:ratio}}
\end{figure}

\begin{figure}
\includegraphics[scale=0.8]{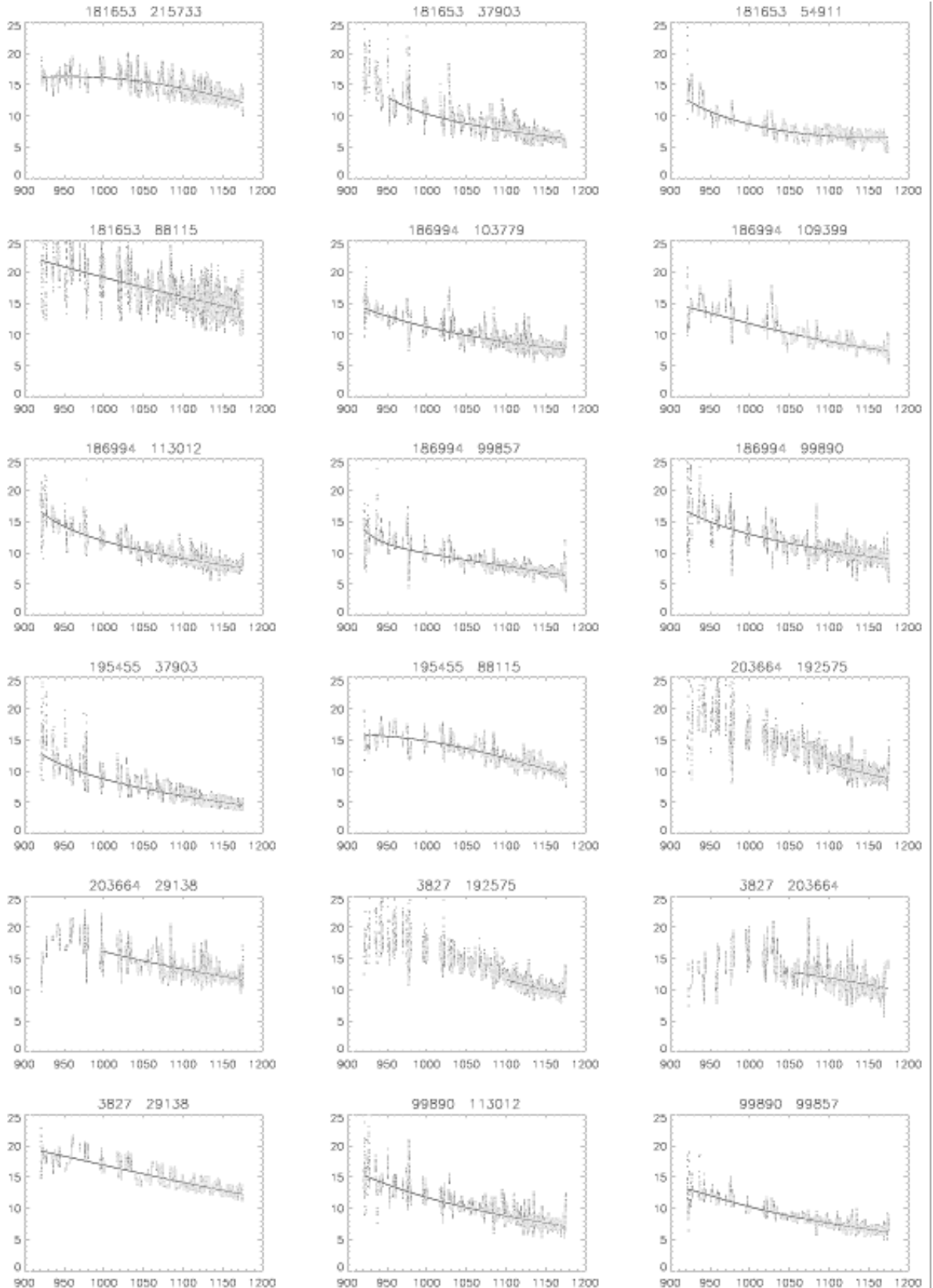}
\caption{The extinction curves ($E(\lambda-V)/E(B-V)$ vs. wavelength in \AA~) derived from 
18 individual star pairs. 
The solid line is the smooth extinction curve derived from these data over
the valid range.  
\label{fig:individual}}
\end{figure}

For each of the star pairs listed in Table~\ref{tab:fits}, 
we first correct the lightly-reddened star using the 
extinction curve of Cardelli, Clayton \& Mathis (1989) extrapolated to 920 \AA\/ 
by fitting second order polynomial to their curve between 1250 and 1700 \AA\/.  
(The Cardelli, Clayton \& Mathis curve is formulated to agree with the Savage \& Mathis
(1979) curve for R$_v = 3.1.$) 
We then divide the two stellar spectra, discarding obviously outlying points 
and perform a least-squares fit to the ratio with a polynomial of 
order 2 -- 4. The exact order selected by eye to provide a good 
characterization of the flux ratio.   We then use this curve to arrive at 
an extinction curve for the pair via Eqn 1.  
In doing so, we apply a normalization factor of 0.9 to the spectra 
taken during ORFEUS I,  on the recommendation of Hurwitz et al. (1998)
for flux agreement with \setcounter{footnote}{0} ORFEUS II\footnote{Flux normalization has since been
standardized for ORFEUS archival data, so this step is not necessary for
data retrieved from the Multimission Archive at STSci.}. 
The data, flux ratio and fit for the pair of stars 
HD 186994 and HD 113012 are shown 
in Figs.~\ref{fig:spec} and~\ref{fig:ratio}.  The gaps in 
the data in Fig.~\ref{fig:ratio} are regions
of the spectrum excluded because of H$_2$ absorption, but the continuum
is still well characterized down to 920 \AA.  
We performed each fit individually and present the range of
validity in Table~\ref{tab:fits}.
The individual extinction curves and the curve fit in the valid range are shown 
in Fig.~\ref{fig:individual}.  

\section{DISCUSSION}

\begin{figure}
\plotone{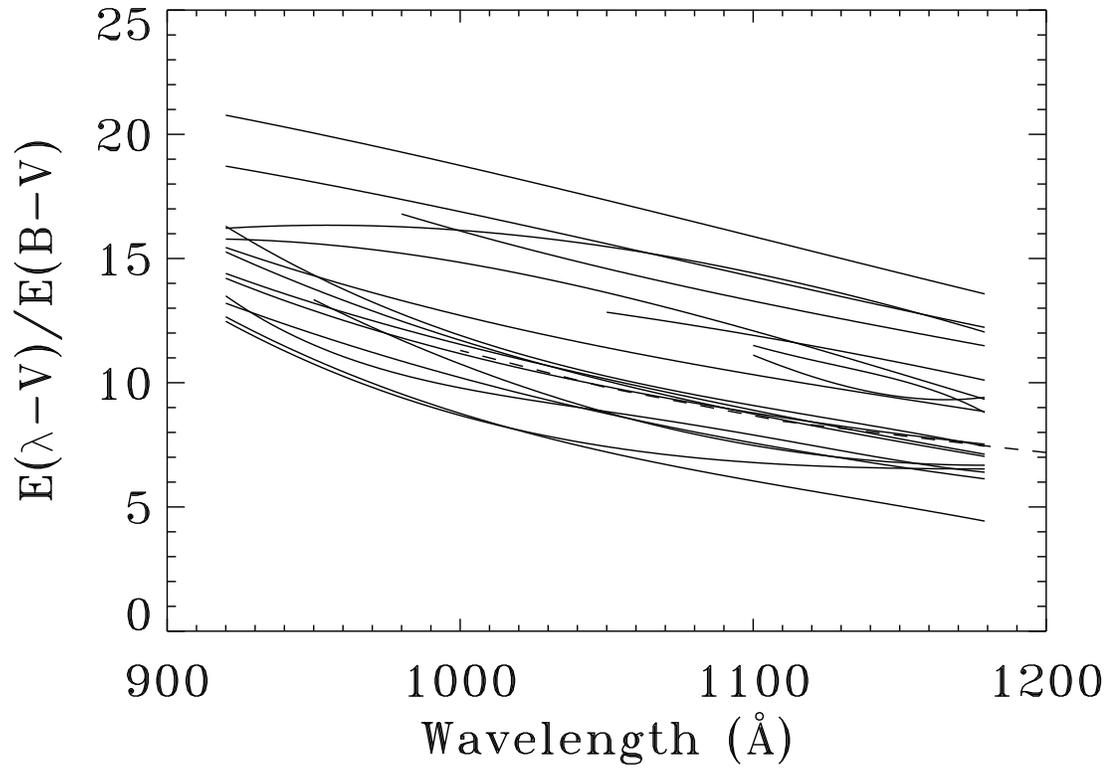}
\caption{The extinction curves derived from the star pairs 
shown in Table 2. The dashed curve is from Savage \&
Mathis (1979).  The scatter in these curves is due both to measurement
uncertainties and intrinsic variations in the extinction.  
\label{fig:all}}
\end{figure}

We show in Fig.~\ref{fig:all} the extinction curves derived from all 18 star pairs, 
together with an extrapolation of the extinction curve 
(dashed line) of Savage \& Mathis (1979).  The extrapolation of the 
Savage \& Mathis curve is done by a second order polymomial fit to 
their standard curve between 2000 and 1000 \AA\/. 
The measured curves show significant scatter and we 
investigate here whether these differences arise from real variations
in the extinction or from measurement uncertainties.  

\subsection{Uncertainties in the Measurement}

A number of potential sources for
uncertainty are discussed by Massa, Savage \& Fitzpatrick (1983)
and Cardelli, Sembach \& Mathis (1992), who find that the three main sources
of systematic error are stellar mismatch, the effects of
an undiscovered stellar companion, and an improper deredding of
the standard star.  
The formal uncertainties for each of the curves 
shown in Fig.~\ref{fig:all} depend on the uncertainties in the quantities in Eqn. 1, which
depend in part on these sources of systematic error. 
By evaluating the size of the uncertainties for the quantities
in Eqn. 1, we can then estimate the uncertainty
for a particular extinction curve.  We use this information to determine
the reliability of a particular curve and then decide whether or not to 
include it in our average.  
This procedure does not specifically evaluate wavelength-dependent effects
in the uncertainties or analysis, as discussed by Massa, Savage \& Fitzpatrick (1983).
However, with the exception of the standard star reddening correction discussed
below, the effect to the average curve from 
uncertainties with random sign is that these uncertainties become 
random and their effect on the average curve is estimated accordingly.  

We have reasonable confidence in our stellar spectral classifications;
the uncertainties of 0.3 in spectral type and half a luminosity class primarly
enter into the uncertainty in $E(B-V)$. 
We estimate the likelihood of significant flux contamination 
from an unknown stellar companion to be small at these wavelengths; the 
companion would have to be of the rare early-B type or earlier to 
affect significantly the total flux from our sample stars. 

An expression for the total uncertainty in an extinction curve is derived
in the appendix.  We use this formula to estimate the uncertainty
for each extinction curve.  We find that even this first-order estimate
of the uncertainty provides a useful measure of the quality of the
extinction curve derived from a given star pair. 
The uncertainties in $V$ and $B-V$, taken from the GCPD, average 
0.014 and 0.009, respectively, for our program stars. 
The uncertainty in $E(B-V)$ is calculated to be 0.022, based on these 
uncertainties and the uncertainty in our spectral identifications. 
We show in Table~\ref{tab:uncert} our estimates of the 
uncertainties in each quantity and their average contribution to the 
overall uncertainty in $E(\lambda-V)/E(B-V)$.  
The latter value is the difference between calculating the uncertainty
(the average of all stars) normally and that found when setting the 
uncertainty in the given parameter to zero. 
The most significant of the individual uncertainties are the elements 
of $E(B-V)$, namely $V$, $(B-V)$ (both taken from GCPD) and $(B-V)_0$, 
based on our spectral type and luminosity
class determinations.  The overall uncertainty for a given curves
varies with the uncertainty in $E(B-V)$.  
Minor variations in $R_v$ are insignificant, as shown in Table~\ref{tab:uncert}.  

\begin{figure}
\plotone{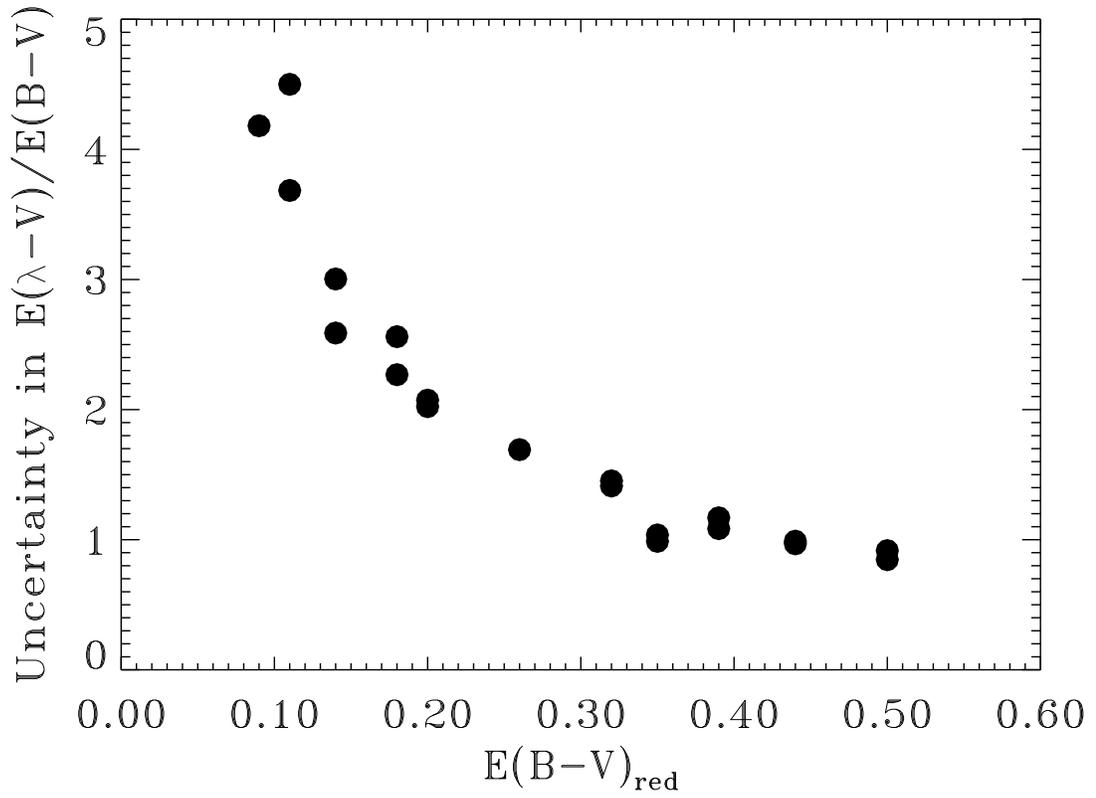}
\caption{The uncertainty in the extinction curves
decreases when more heavily reddened stars are used, with
a much weaker dependence on minor variations in the reddening
of the standard star.  
\label{fig:ebvred}}  
\end{figure}

The largest contribution to the uncertainty derives from the uncertainty
in $E(B-V)$ for the standard star and next from the reddened star.  
The dependence on the standard star reddening arises from the large
reddening correction that must be applied at these wavelengths, while
that from the reddened star directly affects the normalization of
the extinction curve.  
We find that the uncertainty in the extinction curved derived for
a pair of stars is anti-correlated with the reddening  
of the reddened star as shown in Fig.~\ref{fig:ebvred}. This can be understood
because the uncertainties in the stellar magnitudes are essentially independent 
of the reddening and are similar for each star.  Thus the
fractional uncertainty in the reddening of a star
is greatest at low reddening, leading to a larger uncertainty in
the final extinction curve.  
The effect of the $E(B-V)$  uncertainty is most significant
in the highly-extincted FUV wavelengths
measured here and represents a fundamental limitation on the 
use of the pair method at FUV wavelengths.  An accurate measurement
requires a well-reddened star, which in turn implies
faint fluxes in the FUV region, a challenge for current instrumentation. 
At the longer wavelengths studied by Massa, Savage \& Fitzpatrick (1983) 
using IUE data, it was possible to measure reddened stars having
$E(B-V)$ as high as 1.21, leading to lower overall uncertainties.   

The uncertainties we calculate for the individual extinction curves 
derived from each pair (for reference wavelength 1050 \AA) are shown in 
Table~\ref{tab:fits}.  
We show in Figure~\ref{fig:reduced} those curves with 
uncertainty in $E(\lambda - V)/E(B-V)$ less than 1.7 and 
note that the variance is much 
reduced and the average appears close to the Savage \& Mathis (1979)
curve.  It appears critical, then, that an evaluation of the uncertainty
in an extinction curve be made before an assessment of whether the
extinction curve appears anomalously high or low.  
Extinction curves in our sample derived from pairs of stars with reddening 
similar to those of Green et al. (1992) have uncertainties of about 1 calculated
using our $E(B-V)$ uncertainty of 0.022 mag, but the
uncertainty grows to greater than 20 mag in our error formulation 
if we match the $E(B-V)$ uncertainty of 0.5 magnitudes suggested in that paper.  
The anomalously low extinction of $\rho$ Oph is, however, confirmed
by Fitzpatrick and Massa (1990).  
This caution is also relevant to an interpretation of the results of
Buss et al. (1994), who in addition to the uncertainties 
discussed above must incur additional uncertainties 
by correcting for spectral type mismatch and comparing standard
and reddened stars observed with different instruments. 
An excellent summary and discussion of peculiar extinction 
at longer UV wavelengths, including substantial attention to uncertainties,
is given in Massa, Savage \& Fitzpatrick (1983).  

\subsection{The New Measured Average Extinction Curve} 

\begin{figure}
\plotone{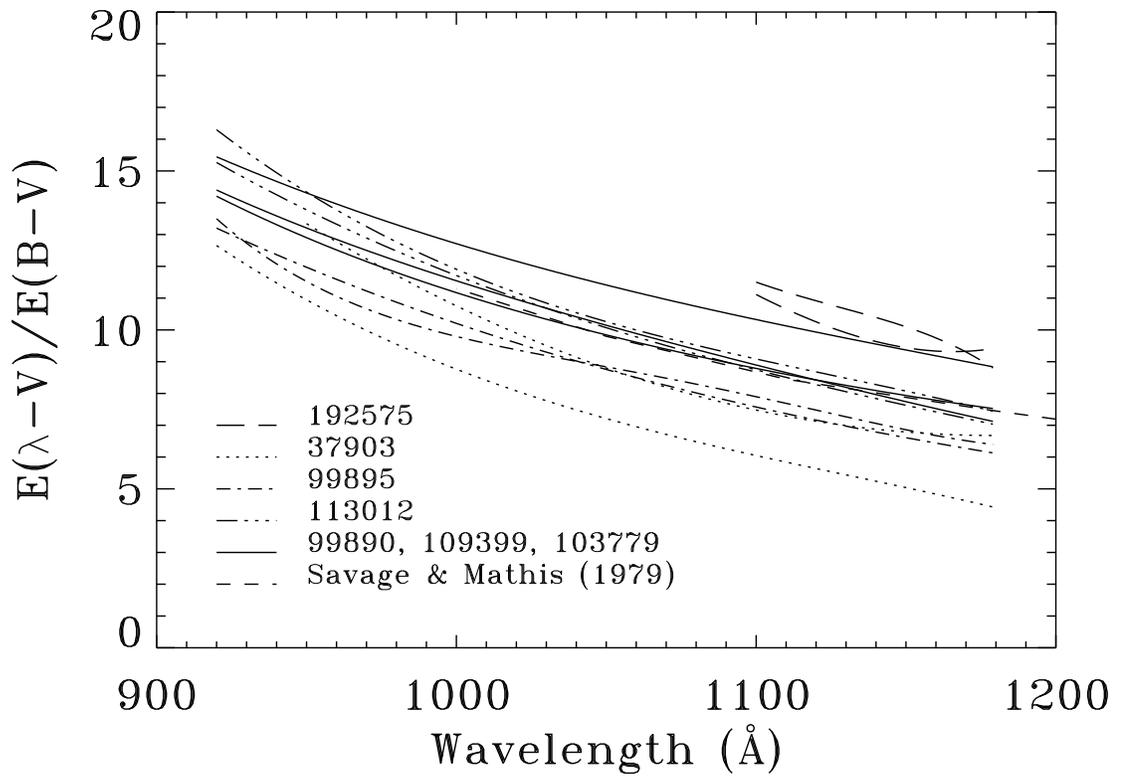}
\caption{The extinction curves derived from the star pairs
with the lowest uncertainties. The dashed curve is from Savage \&
Mathis (1979).  Removing the curves with the highest uncertainty 
substantially reduces the variation in the sample.  
\label{fig:reduced}}  
\end{figure}

\begin{figure}
\plotone{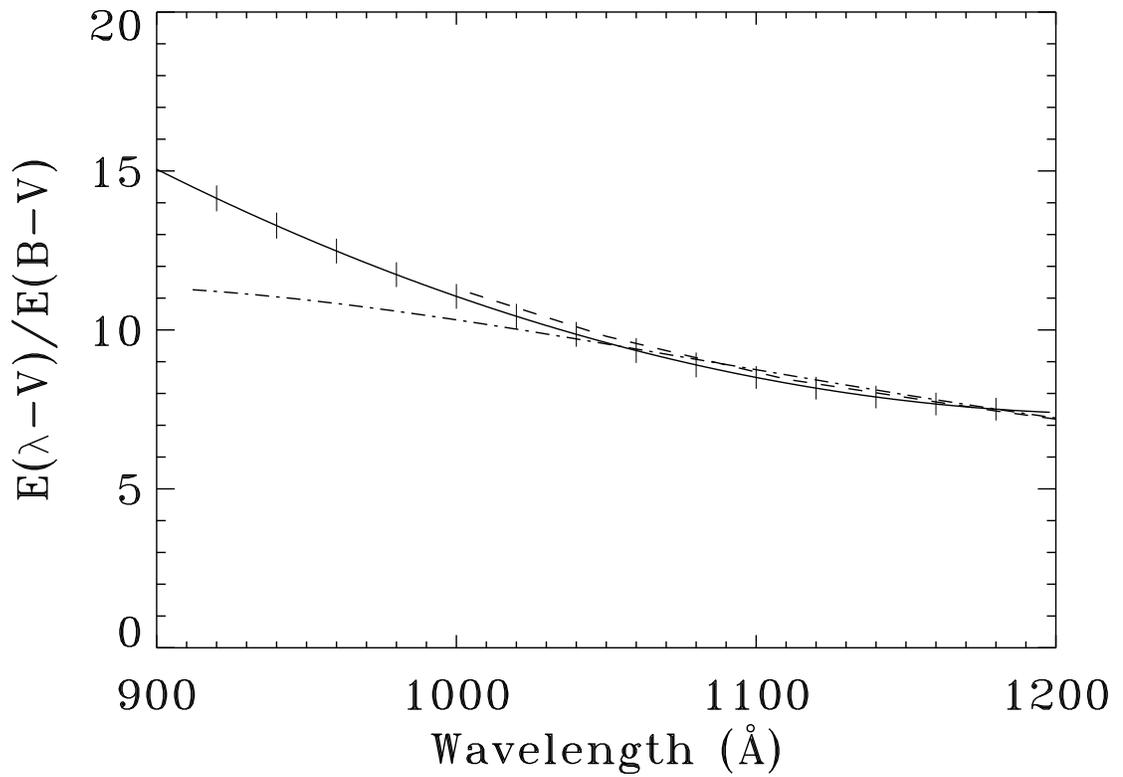}
\caption{The average extinction curve derived from the measurements
with the lowest uncertainties. The dashed curve is from Savage \&
Mathis (1979); the dot-dashed curve is the Draine \& Lee (1984, 1987) 
model discussed in the text. The error bars are derived from the 
uncertainties of the constituent curves at each wavelength shown.  \label{fig:avg}}  
\end{figure}

We have 
averaged the 11 curves shown in 
Fig.~\ref{fig:reduced} with a weighting inverse to their uncertainty
to produce a mean extinction curve. 
To achieve uniform weighting of the seven reddened stars on
which these curves were based, we also halved the weighting of each
pair of curves that were derived from the same reddened star. 
We found the average curve was insensitive to whether we weighted 
each curve or each star equally, with the results agreeing to within 2\%.  
The curve in Fig.~\ref{fig:avg} represents the mean extinction for
the diffuse interstellar medium over the wavelength range 900 -- 1200 \AA,
with the error bars indicating the uncertainty in the average based
on the uncertainty in the individual measurements.   The error bars range
between 0.43 at the blue end to 0.35 at the red end.  
As many authors have discussed, intrinsic luminosity differences lead 
primarily to vertical displacements in the curve, rather than 
significant changes in its shape.  Therefore, averaging several curves 
should yield a good approximation to the
mean extinction.  Curves that lie significantly outside
of the uncertainty range of this mean curve 
may be said to be anomalous.  Coefficients for a polynomial expression 
describing the mean curve in Figure~\ref{fig:avg} valid over the 
range 910 -- 1200 \AA\/ are given in Table~\ref{tab:coeffs}.  
If wavelength is in Angstroms, use the coefficients in the first column;
if wavelength is in $\mu$m$^{-1}$ use the coefficients in the second column
to calculate 
\begin{equation}
\frac{E(\lambda - V)}{E(B-V)} = d_0 + d_1x + d_2x^2 
\end{equation}

The upturn seen at shorter wavelengths in 
Figs.~\ref{fig:reduced} and~\ref{fig:avg} appears to indicate a real 
steepening of the mean extinction curve below 1000 \AA\/ since it occurs 
coherently over many spectral bins for several stars.   
We show in Fig.~\ref{fig:avg}
the Draine \& Lee (1984, 1987) predictions using a grain-size distribution 
taken from Mathis, Rumpl \& Nordieck (1977).  This model 
agrees well with the curve above 1000 \AA, but falls significantly
below the average curve at shorter wavelengths, a finding also reported 
by Green et al (1992) and discussed therein.  
We note that though the standard star correction applied between 
900 and 1000 \AA\/ is a pure extrapolation from longer wavelengths, 
the extrapolated extinction curve actually agrees very well 
with the final result, validating its use in our analysis.  
If instead the true extinction curve were as flat as the model suggests, 
this would have the effect of flattening the curves we derive.  
However, because the upturn
appears in star pairs that have an extinction correction of less than 
a magnitude at 910 \AA, this effect unable to explain the four-unit
discrepancy between the mean curve and the model.  We therefore
infer that the upturn is real and it is the model that needs 
modification.  More detailed modeling of the grain sizes and 
composition leading to the mean and individual extinction curves will 
appear in Paper II.   
The smooth behavior of the extinction curve between
1100 and 1200 \AA\/ does not seem to indicate an extinction bump like the 
2175 \AA\/ feature, as suggested in Fitzpatrick \& Massa (1988).  

We note that our data processing makes our final result 
highly insensitive to small-scale ($< 40$ \AA ) features in the extinction 
curve, if any are present. 
We also note that the distance to stars in this study ranges
from approximately 0.8 to 6.6 kpc. Thus the dust producing the
extinction comes from a range of environmental conditions
and the net effect is an average of all the dust along the
line of sight.  
We examined the 100$\mu$m dust emission maps of Schlegel et al. (1998) for any
indications of unusual concentrations or peculiar dust absorption
along the sightlines of the stars used in this study.  Stars
HD 97991 and HD 195455 were the only ones that were located in
regions of significant dust emission.   These two stars both 
have very low reddening and are used as standard stars in this study.  
This test raised no concerns about peculiar dust properties along
the sightlines to our reddened stars.   We also note the excellent agreement with
the Cardelli, Clayton \& Mathis (1989) curve for $R_v = 3.1 $ in the region of 
wavelength overlap.  
Hence our average curve may be regarded as
typical for the diffuse ISM where an average value for $R_v = 3.1 $ 
is appropriate.  

To examine how differences in luminosity class between the
standard and reddened star affect the resulting extinction curve,
we derived an average extinction curve after eliminating those
three star pairs with luminosity differences of two or more luminosity classes. 
This resulting curve was slightly higher, by less than 2\% everywhere, 
the change being smaller than the uncertainty shown
for the average curve in Fig.~\ref{fig:avg}.  
Two of the curves lie below the average curve while one lies above it, with
the change to the average curve primarily arising because the lowest curve 
in Fig.~\ref{fig:reduced} was eliminated.   We
are hesitant to draw significant conclusions about the effects of luminosity
differences on the extinction curve in this wavelength range based on a sample
of three pairs, but do note that the effect of their inclusion in the 
average curve is not large.  
Cardelli, Sembach \& Mathis (1992) estimate uncertainties resulting from
extreme luminosity-class differences in the wavelength range 1200 - 3200 \AA.
They find that noticeable differences in an extinction curve  
can arise when luminosity mismatch is extreme.  However, they also 
find that this effect, though detectable, does not exceed
the overall uncertainty in the derived extinction curve in that wavelength region.  

\section{CONCLUSIONS}

We have made a new measurement of the FUV dust extinction curve
in the Galactic diffuse interstellar medium from 910 to 1200 \AA,
using high resolution data from the ORFEUS telescope of a carefully selected
sub-sample of B stars and a new model 
for H$_2$ absorption.  This work is the first detailed study of extinction
in this wavelength range using ORFEUS data and the comparatively large sample 
of stars we have available provides a substantially more reliable 
measurement than was previously available.  We find good agreement between
the average of our new measurements and an extrapolation from
longer wavelengths of the standard curve of Savage \& Mathis (1979),
but find considerable individual variations among individual stellar pairs.  
We have shown that measurements of an individual FUV extinction curve
are subject to large uncertainties arising primarily from the $E(B-V)$ 
values adopted for both the standard and reddened star.  
It is clear that credible claims of anomalous extinction curves 
at FUV wavelengths must be accompanied by a careful examination of
the uncertainties in the quantities used to derive them. 

\acknowledgments

The authors would like to thank George Sonneborn for helpful discussions and
useful software. We would like to thank Ken Sembach and an anonymous referee
for helpful comments on 
the manuscript.  We would also like to thank NASA for support
of this research and the ORFEUS program.  This research has made use of the 
SIMBAD database, operated at CDS, Strasbourg, France, and NASA's 
Astrophysics Data System Abstract Service. 

\appendix
\section{FORMULATION OF THE UNCERTAINTY}

The standard-star correction function, $ c(\lambda)$, in Eqn. 1, 

\begin{equation}
E(\lambda - V) = -2.5 \log \left(\frac{F_{red}(\lambda)}{c(\lambda) F_{st}(\lambda)}\right) +
 (V_{st} - E(B-V)_{st}R_{v} - V_{red}) ,
\end{equation}

can be rewritten in terms of $ A_{\lambda} $ for the standard star.  
We calculate $ A_{\lambda,st} $
from 

\begin{equation}
A_{\lambda,st} = E(B-V)_{st} \left(\frac{E(\lambda-V)}{E(B-V)} + R_v\right) .
\end{equation}

The uncertainty in the extinction curve may then be calculated from 
\begin{equation}
\frac{E(\lambda - V)}{E(B-V)} = \frac{1}{E(B-V)_{red}}\left[ -2.5 \log \left(\frac{F_{red}}{F_{st}}\right) + A_{\lambda,st} -E(B-V)_{st}R_v +V_{st} - V_{red} \right] .
\end{equation}

If we define X to be the term in brackets, the uncertainty
in the extinction curve on the left side of Eqn. 5 can be written

\begin{equation}
\delta\left\{ \frac{E(\lambda - V)}{E(B-V)}\right\} = \frac{1}{E(B-V)_{red}}\left[ \left( \frac{X}{E(B-V)_{red}}\right)^2 \sigma_{E(B-V)_{red}}^2 + \sigma_X^2 \right]^{1/2} .
\end{equation}

We can write $\sigma_X$ as 
\begin{equation}
\sigma_X = \{\sigma_\alpha^2 + \sigma_\beta^2 + \sigma_\delta^2 + \sigma_{V_{st}}^2 + \sigma_{V_{red}}^2\}^{1/2} ,
\end{equation}

where
\begin{equation}
\sigma_{\alpha} = 2.5\left[ \left(\frac{\log_{10} e}{F_{st}}\right)^2 \sigma_{F_{st}}^2 + \left(\frac{\log_{10} e}{F_{red}}\right)^2 \sigma_{F_{red}}^2\right]^{1/2} ,
\end{equation}

\begin{equation}
\sigma_{\beta} = \left\{\left( \frac{E(\lambda - V)}{E(B-V)} + R_v\right)^2\sigma_{E(B-V)_{st}}^2 + \left(\sigma_{R_v}^2 + \sigma_{\frac{E(\lambda - V)}{E(B-V)}}^2\right) E(B-V)_{st}^2 \right\}^{1/2} ,
\end{equation}

and 
\begin{equation}
\sigma_{\delta} = \left[E(B-V)_{st}^2\sigma_{R_v}^2 + R_v^2\sigma^2_{E(B-V)_{st}}\right]^{1/2}. 
\end{equation}

We use equation {\em A4} to calculate the uncertainty of an individual extinction curve
at a given wavelength.

\begin{deluxetable} {lcllcclllll}
\small
\rotate
\tablewidth{7.5in}
\tablecaption{BASIC DATA ON PROGRAM STARS AND SPECTRAL TYPE RESULTS\label{tab:basic}}
\startdata
Name & ORF& $l$  & $b$ & V & B-V & \multicolumn{2}{c}{Spectral Type}&$E(B-V)$&$(B-V)_0$& Ref.\\ 
HD  &	1/2 &(deg) &(deg) & (mag) & (mag) & Published & This Work & & & \\ \tableline
3827&	1&	120.79&	-23.23&	8.01&	-0.235&	B0.7Vn& 	B0.5V &	0.05&	-0.28& DS94 \\
29138&	1&	297.99&	-30.54&	7.19&	-0.064&	B1.0Iab& 	B0.2III/&0.20&	-0.27& DS94 \\	
	&	& & & & & &  B0.5Ib-II & & &\\ 
37903&	2&	206.85&	-16.54&	7.83&	0.103&	B1.5V&		B1.5V&	0.35&	-0.25&DS94\\
54911&	1&	229&	-3.06&	7.34&	-0.082&	B2.0II&		B1III&	0.18&	-0.26&DS94\\
88115&	1&	285.32&	-5.53&	8.32&	-0.053&	B1.5IIn&	B1Ib&	0.14&	-0.19&DS94\\
97991&	2&	262.34&	51.73&	7.40&	-0.217&	B2.0V&		B1V&	0.04&	-0.26&DS94\\
99857&	1&	294.78	&-4.94&	7.47&	0.125&	B0.5Ib&		B0II&	0.41&	-0.29&DS94 \\
99890&	1&	291.75	&4.43&	8.30&	-0.059&	B0.0IIIn&	B0III&	0.24&	-0.30&DS94 \\
103779&	2&	296.85&	-1.02&	7.21&	-0.002&	B0.5Iab&	B0III&	0.30&	-0.30&DS94 \\
104705&	1&	297.45&	-0.34&	7.79&	-0.011&	B0Ib&		B0II&	0.28&	-0.29&DS94 \\
109399&	1&	301.71&	-9.88&	7.62&	0.001&	B0.7II&		B0II-III&0.30&	-0.30&DS94 \\
113012&	2&	304.21&	2.77&	8.14&	0.110 &	B0.2Ib&		B0III-IV&0.41& 	-0.30&DS94 \\
121800&	2&	113.01&	49.76&	9.11&	-0.17&	B1.5V&		B2Ib&	0.00&	-0.16&DS94\\
181653&	2&	98.22&	22.49&	8.4&	-0.20&	B1II-III& 	... &	0.05&	-0.25&W71\\
186994&	2&	78.62&	10.06&	7.50&	-0.129&	B0III&		B0III-IV&0.17&	-0.30&HL77\\
192575&	2&	101.44&	18.15&	6.83&	0.166 &	B0.5V &		...  &0.45&	-0.28&C86\\
195455&	1&	20.27&	-32.14&	9.20&	-0.18&	B0.5III&	B1II&	0.06&	-0.24&DS94\\
203664&	2&	61.93&	-27.46&	8.57&	-0.20&	B0.5V &  	...  &	0.08&	-0.28&F94\\
215733&	2&	85.16&	-36.35&	7.33&	-0.135&	B1II&		B1II&	0.11&	-0.24&F94\\
217505&	1&	325.53&	-52.6&	9.14&	-0.208&	B2III&		B2.5IV&	0.00&	-0.22&KBL86\\
219188&	2&	83.03&	-50.17&	7.0&	-0.20 &	B0.5III&	B0.5II&	0.08&	-0.28&F94\\
233622&	1&	168.17&	44.23&	10.01&	-0.21&	B2V&		B2.5II&	0.00&	-0.19&R97\\ 
\enddata
\tablecomments{Mean $V$ and \bv colors are taken from the General Catalog of
Photometric Data by Mermilliod et al. 1997.  Published spectral types key: \\
DS94 -- Diplas \& Savage (1994), 
W71 -- Walborn (1971), 
HL77 -- Hill \& Lynas-Gray (1977),
C86 -- Carnochan (1986), 
F94 -- Fruscione et al. (1994),
KBL86 -- Keenan et al. (1986), 
R97 -- Ryans et al. (1997) }
\end{deluxetable}

\clearpage

\begin{deluxetable}{lllllrcc}
\rotate
\tablewidth{8.3in}
\tablecaption{STAR PAIRS, FIT PARAMETERS, AND UNCERTAINTY\label{tab:fits}}
\startdata
Reddened Star& Standard Star&  \multicolumn{2}{c}{Spectral Type} & $\Delta E(B-V)$&Range Fit & Order of fit & Uncertainty \\ 
HD  &  HD  & Reddened& Standard & & \multicolumn{1}{c}{(\AA)}&  & $\frac{E(\lambda -V)}{E(B-V)}$ \\ \tableline
29138&	3827&   B0.2III/B0.5Ib-II & B0.5V & 0.15 &920-1180& 	3 & 2.02 \\
29138&	203664& B0.2III/B0.5Ib-II & B0.5V & 0.12 &980-1180&	3 & 2.07 \\
37903&	195455& B1.5V & B1II & 0.29 &920-1180&	3 & 0.98 \\
37903&	181653& B1.5V &  B1II-III& 0.30 &1040-1180&	2 & 1.04 \\
54911&	181653& B1III & B1II-III & 0.13 &920-1180&	3 & 2.56 \\
88115&	195455& B1Ib & B1II & 0.08 &920-1180&	3 & 2.59 \\ 
88115&	181653& B1Ib  &  B1II-III & 0.09 &920-1180&	3 & 3.00 \\  
99857&	186994& B0II  & B0III-IV& 0.24 &920-1180&	4 & 0.85 \\
99857&	99890&  B0II  &  B0III & 0.17 &920-1180&	3 & 0.92 \\ 
99890&	186994& B0III  &B0III-IV& 0.07 &920-1180&	3 & 1.69 \\
103779&	186994& B0III & B0III-IV  & 0.13 &920-1180&	4 & 1.45 \\ 
109399&	186994& B0II-III & B0III-IV & 0.13 &920-1180&	4 & 1.41 \\
113012&	99890&  B0III-IV & B0III & 0.17 &920-1180&	3 & 1.17 \\ 
113012&	186994& B0III-IV  &  B0III-IV & 0.24 &920-1180&	3 & 1.09 \\
192575&	3827 &  B0.5V  & B0.5V & 0.40 & 1060-1180& 4 & 0.97  \\ 
192575&	203664& B0.5V  &  B0.5V & 0.37 &1060-1180&	3 & 0.99 \\ 
203664&	3827&   B0.5V & B0.5V & 0.03 & 1050-1180 &	2 & 4.18 \\
215733&	181653& B1II & B1II-III & 0.06 & 920-1180 & 3 & 4.50 \\ 
\enddata
\tablecomments{$\Delta E(B-V)$ is the difference in reddening between the standard and 
reddened star.}
\end{deluxetable}

\clearpage
\begin{table}
\center
\caption{\label{tab:uncert}}
\endcenter
\begin{tabular}{lcc}
\multicolumn{3}{c}{\bf UNCERTAINTIES} \\ \tableline
Quantity &  Typical Uncertainty &  Relative Contribution to Uncertainty\\ 
	& Used	& in $E(\lambda-V)/E(B-V)$ \\ \tableline
$E(B-V)_{st}$: (total, {\em mag}) & 0.02 & 1.0 \\
\hspace{.3in}$(B-V)_o$ & 0.02 & ... \\
\hspace{.3in}B-V &  0.009 & ... \\
$E(B-V)_{red}$: (total, {\em mag}) & 0.022 & 0.75 \\ 
$E(\lambda-V)/E(B-V) $: & 1.0 & 0.1 \\
$F_{st}$ & 5\%  & 0.02 \\	
$F_{red}$ & 5\% & 0.02 \\	
$V_{st}$ {\em mag} & 0.014 & 0.002 \\
$V_{red}$ {\em mag} & 0.014 & 0.002 \\
$R_v$ & 0.2 & 0.008 \\ \tableline
\end{tabular}
\end{table}

\clearpage

\begin{deluxetable}{rrr}
\tablewidth{4in}
\tablecaption{FIT COEFFICIENTS FOR FUV EXTINCTION CURVE\label{tab:coeffs}}
\startdata
$d_i~~~$ & $ x = \lambda $(\AA) & $ x = \lambda^{-1}$  ($\mu m^{-1}$) \\ \tableline
$d_0~~~$ &   116.176  & 34.7847 \\
$d_1~~~$ & -0.177494  &  -7.92908 \\
$d_2~~~$ & 7.23734e-05 &  0.555443 \\ 
\enddata
\end{deluxetable}

\end{document}